# Comparison of different models of non-local impact ionization for low noise avalanche photodiodes

**John S. Marsland** [*]

Department of Electrical Engineering and Electronics, University of Liverpool, Brownlow Hill, Liverpool, L69 3GJ, UK

[*] Corresponding author: e-mail marsland@liv.ac.uk , Phone: +44 151 794 4536, Fax: +44 151 794 4540

The dead space model of non-local impact ionization has been developed independently by three different research groups in order to include a 'soft' dead space. This paper seeks to compare these models for the first time. The models are fitted to Monte Carlo simulation data and it is found that there are no significant differences between the models except that the Liverpool model cannot be used to fit one data set which has a very soft dead space. Likewise the models give very similar results when used to calculate the excess noise factor for low noise avalanche photodiodes.

**1 Introduction** The performance of avalanche photodiodes is known to depend upon the non-local nature of impact ionization. The McIntyre equation [1] for the excess noise factor is derived assuming that the ionization coefficients are a function of the local electric field only. However, when the dead space is considered, the noise is found to be smaller than that predicted by the McIntyre equation. The first non-local theory by Okuto and Crowell [2] has been extended by several researchers to include a so called 'soft' dead space. This paper seeks to compare these various soft dead space theories for the first time.

**2 Hard dead space model**
Okuto and Crowell [2] proposed a model for non-local impact ionization that has subsequently become known as the hard dead space model (HDSM) or the hard-threshold dead-space multiplication theory (HDSMT). The hard dead space model assumes that there is no impact ionization within the dead space and, once the carrier has traversed the dead space, it has a constant non-local ionization coefficient where the non-local ionization coefficient, $\alpha_{OK}(z)$, is defined such that $\alpha_{OK}(z)dz$ is the probability that a carrier, that has not previously ionized, will ionize in the interval $(z, z + dz)$ starting with no kinetic energy at $z = 0$. The Okuto and Crowell ionization coefficient, $\alpha_{OK}(z)$, is given by the following equation for the hard dead space model.

$$\alpha_{OK}(z) = \alpha^* U(z - d) \qquad (1)$$

where $d$ is the dead space length and $\alpha^*$ is the constant ionization rate for carriers that have travelled a distance greater than $d$. $U$ is the unit step function. An ionization pathlength pdf, $h(z)$, can be defined such that $h(z)dz$ is the probability that a carrier ionizes for the first time in the interval $(z, z + dz)$ starting with no kinetic energy at $z = 0$. The following equation gives $h(z)$ for the hard dead space model.

$$h(z) = \alpha^* e^{-\alpha^*(z-d)} U(z - d) \qquad (2)$$

Alternatively the non-local ionization coefficient, $\alpha_{MC}(z)$, can be defined such that $\alpha_{MC}(z)\, dz$ is the probability that a carrier will ionize in the interval $(z, z + dz)$ starting with no kinetic energy at $z = 0$. The carrier can ionize any number of times in travelling to $z$. This definition is often called the Monte Carlo ionization coefficient because it is easy to determine in Monte Carlo calculations e.g. [3]. However its use is not limited to Monte Carlo simulations. $\alpha_{MC}(z)$ can be thought of as the sum of the pdf's for the first ionization, $h(z)$, given by Eq. (2), the second ionization $h_2(z)$, the third $h_3(z)$ and so on. The following equa-tion for $\alpha_{MC}(z)$ has previously been given by [4] for the hard dead space model.

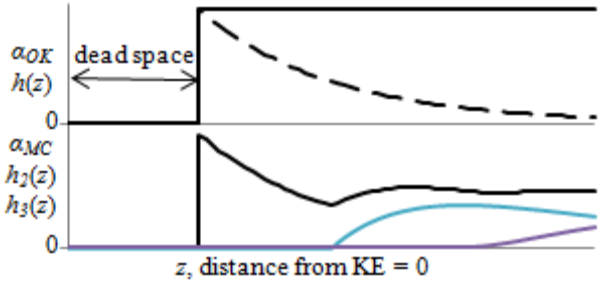

**Figure 1** Non-local ionization coefficients (solid lines) and ionization pathlength pdfs ($h(z)$ dashed line, $h_2(z)$ blue online, $h_3(z)$ mauve online) for the hard dead space model.

$$\alpha_{MC}(z) = \sum_{n=1}^{\infty} h_n(z)$$
$$= \sum_{n=1}^{\infty} \frac{\alpha^{*n}(z-nd)^{n-1}\exp(-\alpha^*(z-nd))U(z-nd)}{(n-1)!} \quad (3)$$

$\alpha_{OK}(z)$, $h(z)$ and $\alpha_{MC}(z)$ are shown diagrammatically in Figure 1.

### 3 Soft dead space model

The abrupt transition at $z = d$ in the non-local ionization coefficient and ionization pathlength pdf, which are a feature of the hard dead space model, are not observed in results obtained by Monte Carlo simulations. This has led to models for a 'soft' dead space to be proposed by three different groups working independently. All models use three parameters and include the hard dead space model as a limiting case when one of those parameters equals zero.

**3.1 Kwon model** The model proposed by Kwon et al [5] starts from an initial assumption that the dead space length can vary uniformly between two limits $d_{min}$ and $d_{max}$. Each individual carrier behaves as described by the hard dead space model and ionizes at a constant rate $\alpha^*$ after it has traversed its own dead space $d$ which is part of a distribution of dead spaces described by a probability density function $f_D(z)$.

$$f_D(z) = \frac{1}{d_{max} - d_{min}} = \frac{1}{\Delta d} \qquad d_{min} < z < d_{max} \quad (4)$$

The ionization pathlength pdf then follows as shown in [5].

$$h(z) = \frac{1}{\Delta d}\left(1 - e^{-\alpha^*(z - d_{min})}\right)U(z - d_{min})$$
$$- \frac{1}{\Delta d}\left(1 - e^{-\alpha^*(z - d_{max})}\right)U(z - d_{max}) \quad (5)$$

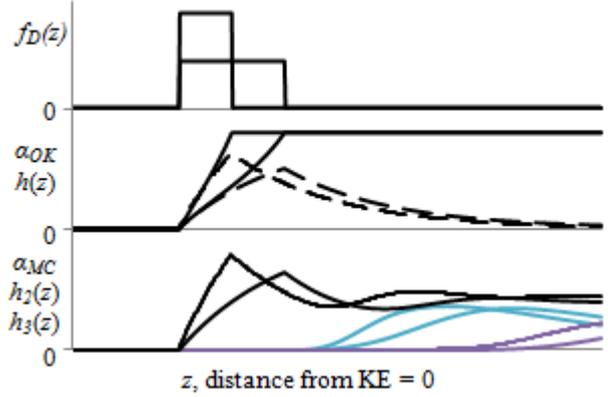

**Figure 2** Non-local ionization coefficients and ionization pathlength pdfs for the Kwon model using two different values of $\Delta d$.

The Okuto and Crowell non-local ionization coefficient, $\alpha_{OK}(z)$, can be calculated as follows.

$$\alpha_{OK}(z) = 0 \qquad z < d_{min}$$
$$\alpha_{OK}(z) = \frac{\alpha^*}{1 + \alpha^*(d_{max} - z)/\left(1 - e^{-\alpha^*(z - d_{min})}\right)} \qquad d_{min} < z < d_{max}$$
$$\alpha_{OK}(z) = \alpha^* \qquad z < d_{max} \quad (6)$$

The Monte Carlo ionization coefficient, $\alpha_{MC}(z)$, is best calculated numerically using the convolution integral given in [6]. The hard dead space model is given when $\Delta d = 0$. The characteristic feature of the Kwon model is the discontinuous derivative of $\alpha_{OK}(z)$, $h(z)$ and $\alpha_{MC}(z)$ at $z = d_{max}$.

**3.2 The Sheffield model** The model proposed by Tan and co-workers from the University of Sheffield [7] starts from the assumption that the Okuto and Crowell non-local ionization coefficient, $\alpha_{OK}(z)$, can be described by one half of a Gaussian curve. They give Eq. (7) where $d$ is the dead space length, $s$ is a spreading factor and $\alpha^*$ is the asymptotic value of $\alpha_{OK}(z)$ in the limit $z \to \infty$.

$$\alpha_{OK}(z) = \alpha^*\left(1 - e^{-(z-d)^2/s^2}\right)U(z - d) \quad (7)$$

This leads to an ionization pathlength pdf given by the following equation.

$$h(z) = \alpha_{OK}(z)\exp\left\{-\alpha^*\left(z - d - \frac{s\sqrt{\pi}}{2}\mathrm{erf}\left(\frac{z-d}{s}\right)\right)\right\} \quad (8)$$

The dead space pdf defined by Kwon et al [5] can be found using the following formula.

$$f_D(z) = h(z) + \frac{1}{\alpha^*} \cdot \frac{dh(z)}{dz} \quad (9)$$

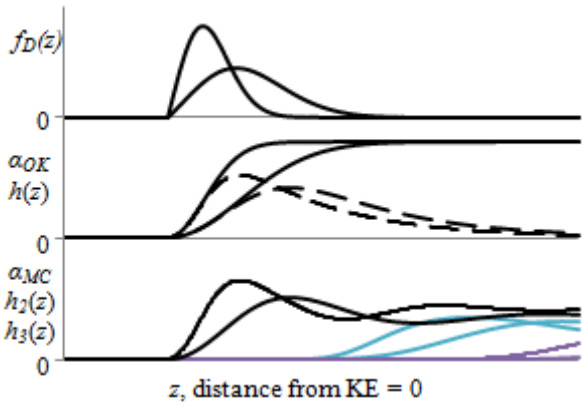

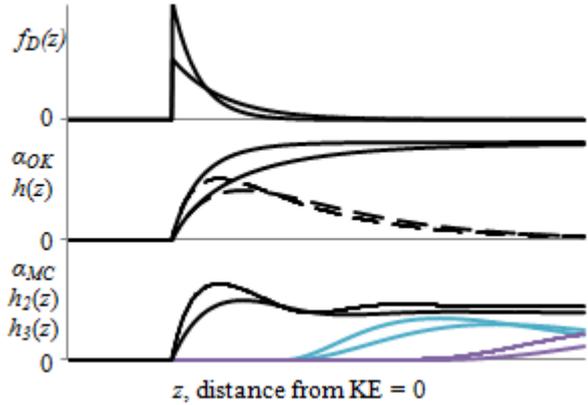

**Figure 3** Non-local ionization coefficients and ionization pathlength pdfs for the Sheffield model using two values of $s$.

**Figure 4** Non-local ionization coefficients and ionization pathlength pdfs for the Liverpool model using two values of $\sigma$.

For the Sheffield model the expression in Eq. (10) describes the dead space pdf.

$$f_D(z) = \alpha^* e^{-(z-d)^2/s^2}\left(1 + \frac{2(z-d)}{\alpha^* s^2} - e^{-(z-d)^2/s^2}\right)$$

$$\times \exp\left\{-\alpha^*\left(z - d - \frac{s\sqrt{\pi}}{2}\mathrm{erf}\left(\frac{z-d}{s}\right)\right)\right\}U(z-d) \quad (10)$$

Again the Monte Carlo ionization coefficient, $\alpha_{MC}(z)$, is best determined by a numerical calculation. The hard dead space model is found by substituting $s = 0$ into the equations above. The slope of $\alpha_{OK}(z)$, $h(z)$ and $\alpha_{MC}(z)$ is zero at $z = d$ and this differentiates the Sheffield model from the other models.

**3.3 The Liverpool model** The soft dead space model proposed by Marsland [6] from the University of Liverpool starts by assuming a shape for the ionization pathlength pdf, $h(z)$. Originally Marsland used three parameters $a$, $b$ and $l$ and introduced an alternative set of parameters $\sigma$, $\lambda$ and $l$ in [4] where $\sigma = \sqrt{(b/a)}$ 'skew' controls the degree of softness and $\lambda = 1/\sqrt{(ab)}$ determines the magnitude of the ionization coefficient. In this paper $d$ is used for the dead space length rather than $l$ for consistency with the other models. The ionization pathlength pdf is assumed to be given by the following equation.

$$h(z) = \frac{\sigma}{\lambda(1-\sigma^2)}\left(e^{-\sigma(z-d)/\lambda} - e^{-(z-d)/\sigma\lambda}\right)U(z-d) \quad (11)$$

Expressions for the non-local ionization coefficients are given in [4] and the dead space pdf can be found using Eq. 9 by substituting $(1/\sigma\lambda)$ for $\alpha^*$.

$$f_D(z) = \frac{1}{\sigma\lambda}e^{-(z-d)/\sigma\lambda}U(z-d) \quad (12)$$

The hard dead space model is obtained when $\sigma = 0$.

**4 Comparison with Monte Carlo data**
The three models described in the previous section have been fitted to seven sets of Monte Carlo data [3, 8, 9]. One data set [3], shown in Fig. 5, is for the non-local ionization coefficient and the other data sets are for $h(z)$. In the case of Fig. 5, the best fit was obtained by the Liverpool model and the worst fit was given by the Kwon model. However the difference between the three models is significantly less than the uncertainty due to the random error that is an intrinsic part of any Monte Carlo calculation. Overall the Sheffield model is best in 3 of the 7 cases and the other two models share the remaining 4 cases, two each. The fitted parameters and ranking are shown in Table 1.

In general there is little to choose between the models for 6 of the 7 data sets. The exception is the data of Ong et al [8] for electrons which reveals a weakness in the Liverpool model. The factor governing the softness of the dead space, $\sigma$, is limited to the range 0 to 1 (equations are equivalent if $\sigma$ is replaced by $1/\sigma$) so that the slope of $h(z)$ at $z = d$ from Monte Carlo data can be smaller than the smallest value allowed by the Liverpool model for $\sigma = 1$.

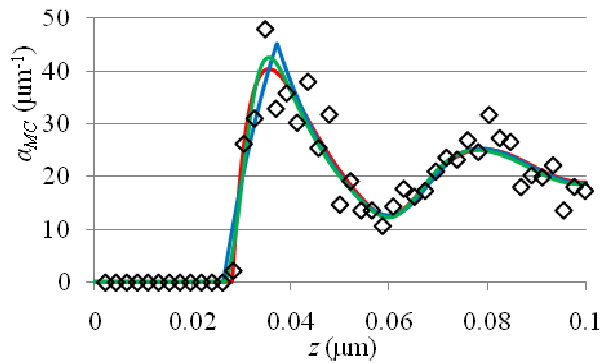

**Figure 5** The three models fitted to the Monte Carlo data of Plimmer et al [3]. Online colour: Kwon model in blue, Sheffield model in green and Liverpool model in red.

**Table 1** Best fit parameters for the three models and ranking for seven sets of Monte Carlo data.

| | Data source: | Plimmer et al [3] | Ong et al [8] | | Rees and David [9] | | | |
|---|---|---|---|---|---|---|---|---|
| | Field: | 70 V$\mu$m$^{-1}$ | 96 V$\mu$m$^{-1}$ | | 90 V$\mu$m$^{-1}$ | | 60 V$\mu$m$^{-1}$ | |
| | Carrier: | E | E | H | E | H | E | H |
| Kwon model | $\alpha^*$ ($\mu$m$^{-1}$) | 62.6 | 103 | 80.7 | 43.1 | 28.5 | 17.2 | 9.87 |
| | $\Delta d$ ($\mu$m) | 0.0109 | 0.0128 | 0.00989 | 0.00867 | 0.0178 | 0.0339 | 0.0377 |
| | $d_{min}$ ($\mu$m) | 0.0263 | 0.0161 | 0.0120 | 0.0412 | 0.0339 | 0.0513 | 0.0388 |
| | rank | 3 | 2 | 3 | 2 | 2 | 1 | 1 |
| Sheffield model | $\alpha^*$ ($\mu$m$^{-1}$) | 60.0 | 126 | 82.6 | 43.7 | 31.1 | 17.6 | 10.1 |
| | $s$ ($\mu$m) | 0.00608 | 0.0130 | 0.00717 | 0.0126 | 0.0208 | 0.0263 | 0.0301 |
| | $d$ ($\mu$m) | 0.0263 | 0.0138 | 0.0113 | 0.0360 | 0.0264 | 0.0468 | 0.0333 |
| | rank | 2 | 1 | 1 | 3 | 1 | 2 | 2 |
| Liverpool model | $\lambda$ ($\mu$m) | 0.00799 | 0.00726 | 0.00700 | 0.00794 | 0.0115 | 0.0302 | 0.0332 |
| | $\sigma$ | 0.532 | 1.000 | 0.646 | 0.342 | 0.312 | 0.577 | 0.327 |
| | $d$ ($\mu$m) | 0.0281 | 0.0184 | 0.0138 | 0.0431 | 0.0414 | 0.0546 | 0.0506 |
| | rank | 1 | 3 | 2 | 1 | 3 | 3 | 3 |

The three models give different asymptotic values for $\alpha_{MC}(z)$ and $\alpha_{OK}(z)$ as $z\rightarrow\infty$. For $\alpha_{OK}(z)$ the differences range from 1.7% to 28% with an average of 11.6% over the 7 data sets whereas for $\alpha_{MC}(z)$ the range is from 0.65% to 7.5% with an average of 2.7%. The greater consistency for $\alpha_{MC}(z\rightarrow\infty)$ compared to $\alpha_{OK}(z\rightarrow\infty)$ can be explained by observing that $\alpha_{OK}(z\rightarrow\infty)$ is determined by the tail of $h(z)$ whereas $\alpha_{MC}(z\rightarrow\infty)$ depends upon the whole distribution.

## 5 Significance for excess noise factor

The excess noise factor, $F(M)$, has been calculated using the methodology described in [10] for the three models using the parameters fitted to the Monte Carlo data of Rees and David [9] for holes at 60 V$\mu$m$^{-1}$ (upper set of curves in Fig. 6) and Ong et al [8] for electrons (lower set of curves). Single carrier multiplication is assumed and the multiplication, $M$, is varied by changing the multiplication region width. Also shown on Fig. 6 are the upper limit given by McIntyres equation [1] and the lower deterministic limit [4, 10]. The difference between the three models is insignificant even for the data of Ong et al [8] where the Liverpool model does not give as good a fit as the other two models.

## 6 Conclusions

Three different models for a soft dead space have been compared and are found to give very similar results when fitted to Monte Carlo data. The differences between the three models are significantly less than random error due to the Monte Carlo calculation. When used to calculate excess noise factor for low noise avalanche photodiodes, the three models show very good agreement.

**Acknowledgements** This paper is dedicated to the memory of my PhD supervisor Peter Robson who introduced me to the study of APDs and encouraged my early work on the dead space effect in 1988 in Sheffield.

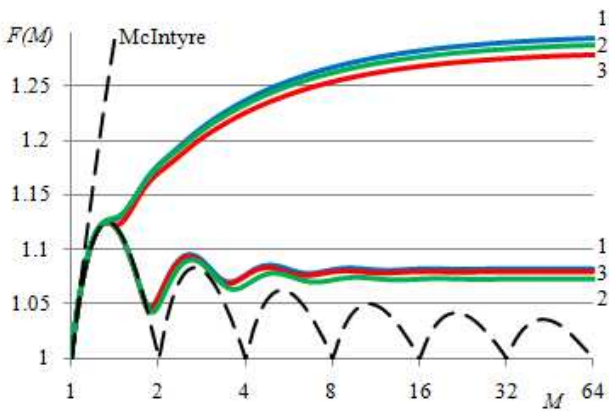

**Figure 6** Excess noise factor calculated for two data sets from Table 1 for all three models; 1 for Kwon model (blue online), 2 for Sheffield model (green) and 3 for Liverpool model (red).